\documentclass[11pt,a4paper]{article}
\usepackage[english]{babel}
\usepackage{amsmath,amssymb,cite}%
\usepackage{hyperref}
\usepackage[large,nohug,heads=vee]{diagrams}
\diagramstyle[labelstyle=\scriptstyle]
\def\eq#1{\begin{equation}#1\end{equation}}
\def\eqs#1{\begin{eqnarray}#1\end{eqnarray}}

\font\Sets=msbm10

\def\Z {\hbox{\Sets Z}}

\newtheorem{thm}{Theorem}[section]

\newtheorem{defn}{Definition}[section]

\newcounter{transf}
\newcounter{meq}
\title{\bf Non-invertible transformations for the classification of differential-difference equations}
\author{{\bf R.N. Garifullin$^{1}$, R.I. Yamilov$^1$ and D. Levi$^2$}
\\$^1$ Ufa Institute of Mathematics, Russian Academy 
of Sciences,\\ 112 Chernyshevsky Street, Ufa 450008, Russian Federation
\\$^2$Department of Mathematics and Physics, Roma Tre University \\
and Sezione INFN  {\it Roma Tre},\\
Via della Vasca Navale 84, 00146 Rome, Italy\\
{\sl E-mails: rustem@matem.anrb.ru, RvlYamilov@matem.anrb.ru,}
\\ {\sl decio.levi@roma3.infn.it}}

\begin{document}
\maketitle

\abstract{We discuss  aspects of the theory of non-invertible transformations which enter in the problem of classification of diffe\-ren\-tial-difference equations and, in particular, the notion of Miura type transformation. We introduce the concept of non--Miura type linearizable transformation and we  present  techniques which allow one to construct simple linearizable transformations and help us to solve the classification problem. This theory is illustrated by the example of a new integrable differential--difference equation depending on 5 lattice points, interesting from the viewpoint of the non-invertible transformation which relate it to an Itoh--Narita--Bogoyavlensky equation.}

\section{Introduction}
\indent The generalized symmetry method uses the existence of generalized symmetries as an integrability criterion and allows one to classify integrable equations of a certain class. Using this method, the classification problem has been solved for some important classes of Partial Differential Equations  (PDEs) \cite{msy87,mss91}, of differential-difference equations \cite{asy00,y06}, and of Partial Difference Equations (P$\Delta$Es) \cite{ly11,gy12}. Classification is carried out in two steps: at first we find all integrable equations of a certain class up to invertible transformations, usually point ones, then we search for non-invertible transformations which relate different resulting equations. For this reason, a theory of non-invertible transformations is necessary. 

This is not the only integrability criterion introduced to produce integrable partial difference equations. Using the Compatibility Around the Cube (CAC) technique introduced by Nijhoff and Walker \cite{n} Adler, Bobenko and Suris \cite{ABS} obtained  a class of integrable equations on a quad graph. More results on this line of research can be found in \cite{abs2,boll}.

Let us consider  autonomouos differential-difference equations of the form:
\eq{\dot u_n=f(u_{n+k},u_{n+k-1},\ldots,u_{n+m}),\ \ k>m,\label{wide}}
where $n\in\Z$ is a discrete variable, $u_n=u_n(t)$ is the unknown function, $\dot u_n$ is its  derivative with respect to the continuous variable $t$.
In \eqref{wide} we can find integrable equations of  Volterra type, corresponding to the case $k=-m=1$. They have been well-studied and a complete list of such equations has been obtained, see e.g. the review article \cite{y06}. In other cases only some integrable examples are known \cite{b91,mx13,hls,1,bo88,ap14,gy12}.

Let us consider the existence of explicit in one direction non-invertible transformations of the form:
\eq{v_{n}=\phi(u_{n+q},u_{n+q-1},\ldots,u_{n+s}),\ \ +\infty>q>s>-\infty,\label{transf}}  
which relate two equations of the form \eqref{wide}.
An example of such a transformation is provided by the following well-known relation  \cite{vv}
\eq{v_n=(1+u_n)(1-u_{n+1}),\label{dM}} which transforms the modified Volterra equation \eq{\dot u_n=(1-u_n^2)(u_{n+1}-u_{n-1})\label{mvolt}} into the Volterra one
\eq{\dot v_n=v_n(v_{n+1}-v_{n-1}).\label{volt}}
This is a discrete analogue of the Miura transformation \cite{m68}
\eq{v=u_x-u^2\label{miura},} 
which relates the Kortevege-de Vries (KdV) equation to the modified KdV one:
$$u_t=u_{xxx}-6u^2u_x,\quad v_t=v_{xxx}+6vv_x, $$ where the indices $t$ and $x$ denote $t$- and $x$-derivatives. The Miura transformation \eqref{miura} is locally non--invertible. 

Differential substitutions in the class $v=\phi(u,u_x)$ of Miura type for evolutionary scalar PDEs have been classified by Startsev \cite{st98} and are, up to point transformations 
\eqs{v=u_x,\qquad v=u_x+u^2, \qquad v=u_x+e^u, \qquad v=u_x+e^u+e^{-u},\label{m1}}  
see also \cite{ssy83,y93}.
The first of the differential substitutions presented by Startsev  in \eqref{m1} is in effect a potentiation, locally non--invertible, but whose inverse is given by an integration. Transformations involving potentiations and point transformations can be very involved and sometimes difficult to distinguish from Miura transformations. 

The notion of Miura type transformation on the lattice is more complicate, not yet well understood, and it is difficult to decide whether a given transformation is of Miura type. Sometimes in the literature  complicate transformations \eqref{transf} are  called of Miura type even if their inversion requires just discrete potentiations.
On the other hand, there are simple non-invertible transformations which look  of Miura type.
  
To be able to define  a Miura type transformation on the lattice we introduce and discuss the rather wide concept of {\it linearizable transformation} not of Miura type. We also present some  techniques  to construct simple linearizable transformations which help us to solve the problem of recognizing Miura type transformations.

In this paper we consider an example which belongs to the class \eqref{wide} with $k=-m=2$,
\eq{\label{p11} \dot u_n =f(u_{n+2},u_{n+1}, u_n, u_{n-1}, u_{n-2}).}Equations of this form \eqref{p11} are currently relevant, as their B\"acklund transformations are completely discrete partial difference equations  \cite{lb,l81,1,mx13,hls,gy12} whose classification is very difficult to perform. Study of  equations \eqref{p11} is in progress, see e.g. \cite{a14,a16,8}.

 Let us consider here the equation
\eq{\label{eq1}\dot u_n = (u_{n+1}-u_n)(u_n-u_{n-1})\left(\frac {u_{n+2}}{u_{n+1}}-\frac{u_{n-2}}{u_{n-1}}\right).}
It has been found by generalized symmetry classification of some particular cases of \eqref{p11}. Currently this classification is in progress, see \cite{8}.
It turns out that \eqref{eq1} is transformed by
\eq{v_n=u_{n+1}+u_{n-1}-u_n-u_{n+1}u_{n-1}/u_n\label{tr_0}}
into the equation
\eq{\label{INB}\dot v_n=v_n(v_{n+2}+v_{n+1}-v_{n-1}-v_{n-2}),}
a well known integrable equation of the Itoh--Narita--Bogoyavlensky class \eqref{p11}  \cite{bo88,i75,na82}.

In Section 2 we will present the notions of Miura type and linearizable transformations and present some  techniques to construct  simple linearizable transformations. Then in Section 3 we will apply our results to the case of \eqref{tr_0}, showing that  it is linearizable and, therefore, not of Miura type. Section 4 is devoted to some concluding remarks. 

\section{Theory}

In this Section  we discuss the notions of Miura type and linearizable transformations and present some  techniques necessary to construct  simple linearizable transformations. 

\subsection{Miura type and linearizable transformations}
The Miura transformation \eqref{miura} is a Riccati equation if we consider  \eqref{miura} as an equation for the unknown function $u$ with $v$ a given function. As it is known, the Riccati equation with $x$--dependent coefficients cannot be solved by quadrature, i.e. by a simple integration. Inversion of the discrete Miura transformation \eqref{dM} is also equivalent to solving the discrete Riccati equation \cite{h79}. From our point of view,  Miura type transformations must be of the same type, i.e. their inversion is somehow reduced to solving a Riccati equation with $x$--dependent coefficients and cannot be done by a simple integration. 

On the other hand, in case of KdV type partial differential equations, we find many transformations  of the form \eq{v=\phi(u,u_x,u_{xx},\ldots),\label{tr_con}} which are a superposition of point transformations $v=\psi(u)$ and of  a potentiation $v=u_x$. In this case finding  $u$ in \eqref{tr_con} is reduced to  integrations. 

In the case of Volterra type equations \eqref{wide} with $k=-m=1$, many transformations of the form of \eqref{transf} are superpositions \cite{y06} of  linear  transformations of the form
 \eq{v_n=u_{n+1}-u_n,\qquad v_n=u_{n+1}+u_n\label{t3}} which are solved by a summation, a {\it discrete potentiation}  and point transformations as $v_n=\psi(u_n)$. Transformations of the form \eq{v_n=u_{n+1}u_n,\qquad v_n=u_{n+1}/u_{n}\label{pr_dev}} are also obviously transformed into equations of the form \eqref{t3}, as $$
v_n=\left(\exp \circ (T\pm 1)\circ \log \right)\,u_n,$$ 
where $T$ is the shift operator $Th_n=h_{n+1}$. 
So \eqref{pr_dev} are solved by a summation and point transformations.
Thus, in these cases, finding the unknown function $u_n$ is reduced to solving a number of linear equations with  constant coefficients, i.e. it  contains  discrete potentiations and point transformations.

Let us pass to the general case of  \eqref{wide} and the non--invertible transformations \eqref{transf}.
We see that such transformations are explicit in one direction. If an equation $A$ is transformed into $B$ by a transformation \eqref{transf}, we will say that this transformation has the {\it direction} from $A$ to $B$ and we will write $A\to B$ as in this direction it is explicit. 

In the general case  \eqref{wide} let us consider the most general form of {\it linear transformations}
\eq{v_n=\nu_k u_{n+k}+\nu_{k-1}u_{n+k-1}+\ldots+\nu_m u_{n+m}+\nu,\ \ k>m, \label{l_tran}}
with constant coefficients. Then we can introduce the following definition:

\begin{defn} \label{d1} A transformation of the form \eqref{transf} is called {\it linearizable} if it can be represented as a superposition of linear transformations \eqref{l_tran} and point transformations $v_n=\psi(u_n)$. In this superposition we allow linear transformations acting in different directions. 
\end{defn}

The linearizable transformation so defined  is decomposed into linear transformations (up to point ones) and thus {\it it is not of Miura type.} 
In an example below we will show in \eqref{pippo} that  a decomposition of  the transformation $A\to B$ of the form \eqref{transf} is possible  $$ A\leftarrow C \rightarrow D\to B,$$ 
and it consists of transformations acting in different directions.

There are two other representations for linearizable transformations which may sometime be useful. For complex constants and functions entering in the transformations  we can represent \eqref{l_tran} as:
$$v_n-\nu=\nu_k(T-\eta_1)(T-\eta_2)\ldots(T-\eta_{k-m+1})T^{m} u_n. $$ 
So, any linearizable transformation is a superposition of elementary transformations of the form:
\eq{\label{kk} v_n=\psi(u_n),\quad v_n=(T-\eta)u_n,\quad v_n=T^{m}u_{n}. }
The second of the transformations presented in \eqref{kk} can be simplified further  as $$v_n=(T-\eta)u_n=\eta^{n+1}(T-1)\eta^{-n}u_n,$$ i.e  it can be reduced to a superposition of transformations of the form:
$$ v_n= \eta u_n,\quad v_n=\alpha^n u_n,\quad v_n=(T-1)u_n. $$ In conclusion any linearizable transformation is reduced, up to a shift,  to a combination of autonomous point transformations, elementary  non--autonomous point transformations and only one concrete non-invertible linear transformation solvable by discrete potentiation.

Let us consider as an example the equation 
\eq{\dot u_n = (u_n^2+au_n)(u_{n+2}u_{n+1}-u_{n-1}u_{n-2}),\label{mikh1}} with $a$ constant. Eq. \eqref{mikh1} is transformed into  \eqref{INB} by the transformation 
\eq{v_n=(u_n+a)u_{n+1}u_{n+2}.\label{to_inb}} When $a=0$ the transformation \eqref{to_inb}  is linearizable, as it is analogous to \eqref{pr_dev}:
$$v_n=\left(\exp\circ (T^2+T+1)\circ \log \right)\,u_n. $$
In the case when $a=0$  \eqref{mikh1} is a well-known modification of \eqref{INB}, see e.g. \cite{b91}. 

The ambiguity in the use of the wording {\it Miura transformation} is present in the many researches in this field. For example in \cite{mx13}  the transformation \eq{v_n=\frac{1}{u_{n+1}u_nu_{n-1}-1} \label{MX}} is called a Miura tranformation. However, up to a point transformation and a shift, \eqref{MX} coincides with  \eqref{to_inb} with $a=0.$

In case $a=1$,  \eqref{mikh1} and the transformation \eqref{to_inb} are considered in \cite{ap14,mx13}. It is shown in \cite{ap14} that the transformation \eqref{to_inb} is an analogue, from the viewpoint of the $L-A$ pair, to \eqref{dM} and therefore it is a transformation  of Miura type. 

There is the following analogy between the discrete transformations (\ref{dM}, \ref{to_inb}) and the Miura transformation \eqref{miura}. It is known that the Riccati equation \eqref{miura} can be rewritten by using the transformation $u=-z_x/z$ as a second order homogeneous linear equation with a nonconstant coefficient $v$: \eq{ z_{xx}+vz=0.\label{2}} In case of \eqref{dM} we introduce $z_n$ by the definition   $u_n=1-z_n/z_{n+1}$ and obtain a discrete linear equation of the same type as \,\eqref{2}:
$$ v_nz_{n+2}-2z_{n+1}+z_n=0.$$
Eq. \eqref{to_inb} with $a=1$ can be written in terms of $z_n,$ by the definition $u_n=z_n/z_{n+1}.$ Eq. \eqref{to_inb} with $a=1$ then takes the form of a third order linear discrete equation:
$$v_nz_{n+3}-z_{n+1}-z_n=0. $$ 
So the transformation \eqref{to_inb} with $a=1$  is not linearizable, in the sense of the Definition \ref{d1}.


\subsection{Linearizable transformations, point symmetries  and conservation laws}
In this subsection we will explain how to construct simple autonomous linearizable, according to the Definition \ref{d1}, transformations of the form \eq{\label{non_inver} v_n=\phi(u_{n+1},u_n)} by using  point simmetries and conservation laws. This result  may be useful in solving classification problems and will allow us to analyze in details in the case of  example  \eqref{eq1}  the transformation \eqref{tr_0}. In the case of KdV type partial differential equations, such a theory exists, see e.g. \cite{s88}.  For both partial differential and differential-difference equations a different theory for the construction of non-invertible transformations,  in particular of  Miura type, has been developed in \cite{y90,y93,y94}.

 For an equation of the form \eqref{wide} we can describe all non-autonomouos point symmetries of the form \eq{\label{psym}\partial_\tau u_n=\sigma_n(u_n), \quad \sigma_n(u_n)\neq0, \; \forall n,}  by solving the determining equation: 
 \eq{\sigma_n'(u_n)f=\sum_{j=m}^k \frac{\partial f}{\partial u_{n+j}}\sigma_{n+j}(u_{n+j}),\label{det_eq}} where by a $'$ we mean the derivative of the function with respect to its argument.
If we introduce the  point transformation:
$$ \hat u_n=\eta_n(u_n), \quad \eta'_n(u_n)=\frac1{\sigma_n(u_n)},$$
then the point symmetry \eqref{psym} turns into \eq{ \partial_\tau \hat u_n=1 \label{psym1}} and  \eqref{wide} into an equation of the form:
\eq{\partial_t{\hat u_n}=\hat f_n(\hat u_{n+k},\hat u_{n+k-1},\ldots, \hat u_{n+m}).\label{wide1}}
As \eqref{wide1} admits the symmetry  \eqref{psym1} the determining equation \eqref{det_eq} reduces to \eq{\sum_{j=m}^k \frac{\partial \hat f_n}{\partial \hat u_{n+j}}=0.\label{det_eq1}} This shows that 
 $$ \hat f_n=g_n(\hat u_{n+k}-\hat u_{n+k-1},\hat u_{n+k-1}-\hat u_{n+k-2},\ldots, \hat u_{n+m+1}-\hat u_{n+m}),$$
as \eqref{det_eq1} is a simple first order linear PDE for the function $\hat f_n$ which can be solved on the characteristics.
So we can use the non-invertible linearizable transformation $v_n=\hat u_{n+1}-\hat u_{n},$
to reduce  \eqref{wide} to the equation
\eq{\label{pluto}\dot v_n=(T-1)g_n(v_{n+k-1},v_{n+k-2},\ldots,v_{n+m}).}
We can summarize the previous results in the following Theorem:
\begin{thm} \label{th1}
If  \eqref{wide} has a point symmetry \eqref{psym} with $\sigma_n(u_n)\neq0$ for all $n$, then it admits the following non-invertible and linearizable transformation
 \eq{v_n=(T-1)\eta_n(u_n),\quad \eta'_n(u_n)=\frac1{\sigma_n(u_n)}\label{trans_th}}
 which allows us to construct a modified equation \eqref{pluto}.
\end{thm}

We are mainly interested in autonomous linearizable transformations of the form \eqref{transf}, therefore, primarily in autonomous point symmetries \eqref{psym}. However, sometimes a non-autonomous point symmetry of the form \eqref{psym} may also lead to an autonomous result. This is the case when there exists a non-autonomous point transformation \eq{\label{p26}\tilde v_n=\psi_n(v_n)} which turns the transformation \eqref{trans_th} into an autonomous one . In this case the resulting equation for $\tilde v_n$ turns out to be  also autonomous. 

Indeed, the autonomous equation \eqref{wide} is transformed by the autonomous transformation \eqref{non_inver} into \eqref{pluto}, i.e. 
\eq{\dot v_n=G_n \Big(v_{n+k}, v_{n+k-1}, \ldots , v_{n+m} \Big).}
Differentiating \eqref{non_inver} with respect to the continuous variable $t$ we get 
\eqs{\nonumber &&G_n \Big(\phi(u_{n+k+1},u_{n+k}), \phi(u_{n+k},u_{n+k-1}),\ldots, \phi(u_{n+m+1},u_{n+m}) \Big)=\\ \nonumber &&\qquad \frac{\partial \phi(u_{n+1},u_{n})}{\partial u_n} f(u_{n+k},\ldots,u_{n+m})+\frac{\partial \phi(u_{n+1},u_{n})}{\partial u_{n+1}} f(u_{n+k+1},\ldots, u_{n+m+1}).}
This shows that $G_n$ cannot have any explicit dependence on $n$, i.e. the equation for $v_n$ is autonomous.

Even if \eqref{pluto} is autonomous we can also simplify the result by applying \eqref{p26} with an autonomous function $\psi$.

Let us write down in Table \ref{t1} the four most typical examples of point symmetries, two of which not autonomous, together with corresponding autonomous transformations \eqref{non_inver}. 

\begin{table}[ht]
\begin{center}
\begin{minipage}{0.57\textwidth}
\caption{\small Examples of point symmetries \eqref{psym}, corresponding autonomous linearizable transformations \eqref{non_inver} and point transformations \eqref{p26}.  }
\label{t1}
\end{minipage}
		\begin{tabular}{|c|c|c|c|c|}
			\hline
			$\sigma_n(u_n)$ & $1$ & $(-1)^n$& $u_n$& $(-1)^nu_n$ \\
			\hline
			$\phi(u_{n+1},u_n)$ & $u_{n+1}-u_{n}$& $u_{n+1}+u_{n}$ &$u_{n+1}/u_{n}$&$u_{n+1}u_{n}$\\
			\hline
			$\psi_n(v_n)$ & $v_n$ & $(-1)^{n+1}v_n$& $\exp{v_n}$& $\exp{[(-1)^{n+1}v_n]}$ \\
			\hline
		\end{tabular}
\end{center}
\end{table}

We can construct simple autonomous linearizable transformations also starting from conservation laws. Let us now show how.
From \eqref{wide} we can look for conservation laws of the form \eq{\label{claw}\partial_t\rho_n(u_n)=(T-1)h_n,\quad \rho_n'(u_n)\neq0, \; \forall n,}  where $h_n=h_n(u_{n+k-1},\ldots,u_{n+m})$ with $k > m$. The conserved density $\rho_n$ is found  using the criterion introduced in  \cite{ly97}. A function $\rho_n(u_n)$ is a conserved density  of  \eqref{wide} iff 
\eq{\frac{\delta(\partial_t\rho_n(u_n))}{\delta u_n}=\frac{\delta(\rho_n'(u_n)f)}{\delta u_n}\equiv\sum_{j=m}^kT^{-j}\frac{\partial (\rho_n'(u_n)f)}{\partial u_{n+j}}=0.} 
Let us introduce  a  variable $w_n$, so that
$\rho_n(u_n)=w_{n+1}-w_n.$
It follows from \eqref{claw} that 
\eq{\dot w_n=h_n(\rho_{n+k-1}^{-1}(w_{n+k}-w_{n+k-1}),\ldots,\rho_{n+m}^{-1}(w_{n+m+1}-w_{n+m}))\label{claw1}}
and we can enunciate the following Theorem:
\begin{thm} \label{th2}
If  \eqref{wide} has a conservation law \eqref{claw} with $\rho_n'(u_n)\neq0$ for all $n$, then it admits the following non--autonomous non--invertible linearizable transformation:
 \eq{u_n=\rho_n^{-1}(w_{n+1}-w_n)\label{trans_th2}}
 which allows us to construct a modified equation \eqref{claw1}.
\end{thm}
As  we are looking for  linearizable transformations of the form \eqref{transf} then we need
\eq{\label{non_inver1} u_n=\psi(w_{n+1}, w_n),}
i.e.  we are concerned primarily with autonomous conservation laws. However, sometimes a non-autonomous conservation law may also lead to autonomous result. This is the case when a non-autonomous point transformation \eq{\label{claw2} \tilde w_n=\mu_n(w_n)} makes the transformation \eqref{trans_th2} autonomous, i.e. of the form \eqref{non_inver1}. In this case introducing \eqref{claw2} into \eqref{claw1} and renaming $\tilde w_n$ as $w_n$ to simplify the notation, we get:
\eq{\label{p1}\dot w_n = H_n( w_{n+k}, w_{n+k-1},\ldots, w_{n+m}).}
In this case we can only show  that \eqref{p1} for $ w_n$ is close to an autonomous one as we have:
\eq{\partial_t { w}_n=\tilde H( w_{n+k},\ldots, w_{n+m})+Q_n( w_n).\label{eq_t_w}} In all cases we considered up to now  we can pass to the autonomous equation \eq{\label{eq_t_w_a}\partial_t { w}_n=\tilde H( w_{n+k},\ldots, w_{n+m})} as $\partial_\tau w_n= Q_n( w_n)$ is a point symmetry of   \eqref{eq_t_w}, and \eqref{non_inver1} transforms both \eqref{eq_t_w} and \eqref{eq_t_w_a}  into the same equation \eqref{wide}. 

Let us see the passage from \eqref{p1} to  \eqref{eq_t_w} more in details: let us assume that \eqref{p1} is transformed into \eqref{wide} by the transformation \eqref{non_inver1}. Then we can differentiate \eqref{non_inver1} with respect to the continuous variable $t$ and we get:
\eqs{\label{p2} &&f \Big(\psi(w_{n+k+1}, w_{n+k}), \psi(w_{n+k}, w_{n+k-1}), \ldots, \psi(w_{n+m+1}, w_{n+m})\Big)= \\ \nonumber && \qquad =\frac{\partial \psi(w_{n+1},w_{n})}{\partial w_n}H_n+\frac{\partial \psi(w_{n+1},w_{n})}{\partial w_{n+1}}H_{n+1}.}
If $k>0$ and $m<0$ then we can differentiate with respect to $w_{n+k+1}$ and $w_{n+m}$ and we get:
\eqs{\label{q1} &&\frac{\partial f}{\partial u_{n+k}} T^k \frac{\partial \psi}{\partial w_{n+1}} = \frac{\partial \psi}{\partial w_{n+1}} \frac{\partial H_{n+1}}{\partial w_{n+k+1}},\\ \label{q2}
&&\frac{\partial f}{\partial u_{n+m}} T^m \frac{\partial \psi}{\partial w_{n}} = \frac{\partial \psi}{\partial w_{n}} \frac{\partial H_{n}}{\partial w_{n+m}}.}
From (\ref{q1}, \ref{q2}) we deduce that $\frac{\partial H_{n}}{\partial w_{n+k}}$ and $\frac{\partial H_{n}}{\partial w_{n+m}}$ are autonomous and consequently 
\eq{H_n=\tilde H(w_{n+k}, \ldots, w_{n+m}) + \hat H_n(w_{n+k-1}, \ldots, w_{n+m+1}).\label{qq}}
Substituting \eqref{qq} into  \eqref{p2} we get an expression indicating that $\frac{\partial \psi}{\partial w_n} \hat H_n+\frac{\partial \psi}{\partial w_{n+1}} \hat H_{n+1}$ is also autonomous and we can repeat the procedure.  At the end we get
\eq{H_n=\tilde H(w_{n+k}, \ldots, w_{n+m}) +Q_n(w_n),}
where $Q_n$ is such that $\frac{\partial \psi}{\partial w_n}  Q_n+\frac{\partial \psi}{\partial w_{n+1}} Q_{n+1}=0$. This shows that on the r.h.s. of \eqref{p2} $H_n$ can be replaced by $\tilde H$. Consequently \eqref{p1} and $\dot w_n=\tilde H$ are transformed by the same transformation \eqref{non_inver1} into the same equation \eqref{wide}. 

In the particular case when  \eq{\rho_n(u_n)=(-1)^n p(u_n)\label{den_p}} this procedure is more obvious. In this case we have the representation
\eq{\label{c_law_p}\partial _t p(u_n)=(T+1) \hat h_n, } where $\hat h_n=(-1)^{n+1}h_n$. As the left hand side of \eqref{c_law_p} is autonomous, we can easily prove that $$\hat h_n=\tilde h(u_{n+k-1},\ldots,u_{n+m})+(-1)^nc,\quad c \in \mathbb C,$$ and $(-1)^nc$ is in the null space of $T+1$. So, there is a representation \eqref{c_law_p} with the autonomous function $\tilde h$ instead of $\hat h_n$. In this case the autonomous linearizable transformation and the new equation can be defined as follows: $$p(u_n)=w_{n+1}+w_n,\quad \partial_t w_n=\tilde h.$$
 
 Let us write down in Table \ref{t2} four most typical examples of conserved densities together with the corresponding  autonomous transformations \eqref{non_inver1}. Two of these conserved densities are non-autonomous and have the special form \eqref{den_p}. 

\begin{table}[ht]
\begin{center}
\begin{minipage}{0.59\textwidth}
\caption{\small Examples of conserved densities of \eqref{claw} and  corresponding autonomous linearizable transformations \eqref{non_inver1}, up to autonomous point transformations of $w_n$.}
\label{t2}
\end{minipage}
		\begin{tabular}{|c|c|c|c|c|}
			\hline
			$\rho_n(u_n)$ & $u_n$ & $(-1)^nu_n$& $\log u_n$& $(-1)^n\log u_n$ \\
			\hline
			$\psi(w_{n+1},w_n)$ & $w_{n+1}-w_{n}$& $w_{n+1}+w_{n}$ &$w_{n+1}/w_{n}$&$w_{n+1}w_{n}$\\
			\hline
		\end{tabular}
\end{center}
\end{table}

For a given conserved density $\rho_n(u_n),$ if  the representation \eqref{claw} contains instead of $T-1$ a higher order linear difference  operator with constant coefficients, we can get more linearizable transformations. For instance, for the Volterra equation given in \eqref{volt}, one has the following conservation law
\eq{\partial_t \log v_n=(T^2-1)v_{n-1}=(T-1)(T+1)v_{n-1}.\nonumber} In this case we can construct three  linearizable transformations in a quite similar way:
\eq{v_n=\frac{u_{n+1}}{u_{n}},\quad v_n=u_{n+1}u_n,\quad v_n=\frac{u_{n+2}}{u_n},\label{claw3}}
with the corresponding modified Volterra equations. For instance, from the last of the transformations in  \eqref{claw3} we get the following modified Volterra equation $$\dot u_n=\frac{u_{n+1}u_n}{u_{n-1}}.$$

A further example is provided by  \eqref{INB}. Considering   the  conserved density $\log v_n$ we get:
\eqs{\partial_t \log v_n=(T^4+T^3-T-1)v_{n-2}=(T-1)(T+1)(T-c_1)(T-c_2)v_{n-2},\nonumber} where $c_{1,2}=-\frac12\pm\frac{\sqrt{3}i}{2}$. Here we have more possibilities. If we start from the operator $T-c_1$ instead of $T-1$, we can construct the transformation $v_n=u_{n+1}u_n^{-c_1}.$
Another possibility is provided by the operator $$(T-c_1)(T-c_2)=T^2+T+1$$ which leads to the known transformation \eqref{to_inb} with $a=0$. It is clear that this transformation is the superposition of the following ones: $$v_n=z_{n+1}z_n^{-c_1},\quad z_n=u_{n+1}u_n^{-c_2}.$$

As a final example we consider  \eqref{mikh1}. In this case  one has 
\eq{\partial _t \int\frac{du_n}{u_n^2+au_n}=(T^3-1)u_{n-1}u_{n-2}. \label{claw4}}Eq. \eqref{claw4} provides a number of possibilities, too.

\section{Discussion of  the example  \eqref{eq1}.}

A transformation of the form \eqref{transf} transforming \eqref{eq1} into \eqref{INB} can be found by a straightforward investigation \cite{y06}. We can fix $s=-1$ and $q=1$ in \eqref{transf} and then we find the explicit transform \eqref{tr_0}.

Let us try to find \eqref{tr_0} as a chain of linearizable transformations relating  \eqref{eq1} and \eqref{INB} by applying the  theory presented in Section 2.
Let us  start from  \eqref{eq1} and find a conservation law of  density $\rho_n(u_n)=(-1)^n\log u_n$. 
According to Table \ref{t2} this gives the transformation $u_n=w_{n+1}w_n$ which relates  \eqref{eq1} to
\eq{\label{eq1_1} \dot w_n=(w_{n+2}-w_n)(w_{n+1}-w_{n-1})(w_n-w_{n-2}).}
We can now apply the obvious transformation $z_n=w_{n+1}-w_{n-1}$. This is the superposition of first two transformations of Table \ref{t1} which are provided by the point symmetry $$\partial_\tau w_n=\alpha+\beta(-1)^n. $$ The resulting equation reads:
\eq{\label{eq12} \dot z_n=z_n(z_{n+2}z_{n+1}-z_{n-1}z_{n-2}).}
This is the well-known modification of \eqref{INB}. The transformation of \eqref{eq12} into \eqref{INB} is $v_n=z_{n+1}z_n$ and it corresponds to the symmetry $\partial_\tau z_n=(-1)^nz_n.$

We can construct the same chain of linearizable transformations, moving in the opposite direction, i.e. starting from \eqref{INB}.  
We are led to the following picture:
\eq{\begin{diagram} \label{pippo}
C:\eqref{eq1_1}  & 
\rTo{\eta_2:\ z_n=w_{n+1}-w_{n-1}} & D:\eqref{eq12}
\\
\dTo{\eta_1:\ u_n=w_{n+1}w_n} & & \dTo{}{\eta_3:\ v_n=z_{n+1}z_n} \\ 
 A:\eqref{eq1}& \rTo{\eta:\ \eqref{tr_0}} & B:\eqref{INB}
\end{diagram}}
It turns out that the superposition of the three linearizable transformations shown in the picture
can be rewritten as \eqref{tr_0} which is of the form \eqref{transf}: $\eta=\eta_3\circ\eta_2\circ\eta_1^{-1}$. 

We see that in order to find the unknown function $u_n$ in the non--autonomouos nonlinear discrete equation \eqref{tr_0}, one has to find $w_n$ by solving two linear problems $v_n=z_{n+1}z_{n}$ and $z_n=w_{n+1}-w_{n-1}$ and then to construct $u_n$ using the  explicit formula $u_n=w_{n+1}w_n.$

\section{Conclusions}
In this article we discuss the transformations necessary to classify differential--difference equations. By a Definition \ref{d1} we introduce what we call a {\it linearizable transformation}, e.g. a linear transformation in one direction which in the other direction corresponds to one of the many kinds of discrete potentiations. This is  not a  transformation of  Miura type as its inversion is also simple, being usually of discrete potentiation type. Using this definition we are able to introduce two Theorems, \ref{th1} and \ref{th2} which allow us to construct transformations to a new differential--difference equation, if  the starting one has point symmetries or conservation laws. These two Theorems are used in Section 3 for showing that \eqref{eq1} and \eqref{INB} are related by a linearizable transformation. 

This result is the starting point for a subsequent work \cite{8}  in progress on the classification of a class of differential--difference equations of the form \eqref{wide} with $k=2$ and $m=-2$ which extends the classification of Volterra type equations, corresponding to $k=1$ and $m=-1$, carried out with great success in \cite{y83}.

\paragraph{Acknowledgments.}  RIY has been partly supported by the Russian Foundation for Basic Research
(grant no. 14-01-97008-r-povolzhie-a). DL has been partly supported by the Italian Ministry of Education and Research, 2010 PRIN {\it Continuous and discrete nonlinear integrable evolutions: from water waves to symplectic maps} and   by INFN   IS-CSN4 {\it Mathematical Methods of Nonlinear Physics}.

\end{document}